\documentstyle[manuscript,eqsecnum,aps,version2]{revtex}
\draft
\begin{document}

\preprint{WUHEP-95-17}

\title
\bf Spherically Symmetric Random Walks I. \\
\bf Representation in Terms of Orthogonal Polynomials
\endtitle

\author{Carl M. Bender}
\instit
Department of Physics, Washington University, St. Louis, MO 63130
\endinstit
\author{Fred Cooper}
\instit
Theoretical Division, Los Alamos National Laboratory, Los Alamos, NM 87545
\endinstit
\author{Peter N. Meisinger}
\instit
Department of Physics, Washington University, St. Louis, MO 63130
\endinstit

\medskip
\centerline{\today}

\abstract
Spherically symmetric random walks in arbitrary dimension $D$ can be described
in terms of Gegenbauer (ultraspherical) polynomials. For example, Legendre
polynomials can be used to represent the special case of two-dimensional
spherically symmetric random walks. In general, there is a connection between
orthogonal polynomials and semibounded one-dimensional random walks; such a
random walk can be viewed as taking place on the set of integers $n$,
$n=0,~1,~2,~\ldots$, that index the polynomials. This connection allows one to
express random-walk probabilities as weighted inner products of the polynomials.
The correspondence between polynomials and random walks is exploited here to
construct and analyze spherically symmetric random walks in $D$-dimensional
space, where $D$ is {\sl not} restricted to be an integer. The weighted
inner-product representation is used to calculate exact closed-form spatial and
temporal moments of the probability distribution associated with the random
walk. The polynomial representation of spherically symmetric random walks is
also used to calculate the two-point Green's function for a rotationally
symmetric free scalar quantum field theory.
\endabstract


\section{INTRODUCTION}
\label{s1}
In two recent papers \cite{r1,r2} we proposed and analyzed a new kind of
$D$-dimensional random walk that is well defined even when $D$ is noninteger.
This random walk takes place on a spherical lattice consisting of an infinite
set of concentric nested spheres of radii $R_n$, $n=1,~2,~3,~\ldots~$. We define
$region~n$ to be the volume lying between $R_{n-1}$ and $R_n$, with the central
region, $region~1$ being the volume inside $R_1$. If the random walker occupies
$region~n$ at time $t$, then at time $t+1$ the random walker must move out to
$region~n+1$ with probability $P_{\rm out}(n)$ or in to $region~n-1$ with
probability $P_{\rm in}(n)$. The probabilities of moving out and in are in
proportion to the hyperspherical surface areas bounding $region~n$. Let $S_D(R)$
represent the surface area of a $D$-dimensional hypersphere
\begin{eqnarray}
S_D(R)={{2\pi^{D/2}}\over{\Gamma(D/2)}} R^{D-1}.
\nonumber
%\label{e1.1}
\end{eqnarray}
We then take (for $n>1$)
\begin{eqnarray}
P_{\rm out}(n) = {{S_D(R_n)}\over{S_D(R_n)+S_D(R_{n-1})}}=
{{R_n^{D-1}}\over{R_n^{D-1}+R_{n-1}^{D-1}}}
\label{e1.2a}
\end{eqnarray}
and
\begin{eqnarray}
P_{\rm in}(n) = {{S_D(R_{n-1})}\over{S_D(R_n)+S_D(R_{n-1})}}=
{{R_{n-1}^{D-1}}\over{R_n^{D-1}+R_{n-1}^{D-1}}}.
\label{e1.2b}
\end{eqnarray}
For the special case $n=1$ we define
\begin{eqnarray}
P_{\rm out}(1)=1, \quad P_{\rm in}(1) = 0.
\label{e1.2c}
\end{eqnarray}
Note that probability is conserved because the total probability of the random
walker moving out or in is {\sl unity}:
\begin{eqnarray}
P_{\rm out}(n) + P_{\rm in}(n) = 1.
\label{e1.3}
\end{eqnarray}

To describe a random walk on this lattice we introduce the notation $C_{n,t;m}$,
which represents the probability that a random walker, initially in $region~m$
at time $t=0$, will be found in $region~n$ at time $t$. The probability
$C_{n,t;m}$ satisfies the partial difference equation
\begin{eqnarray}
C_{n,t;m} = P_{\rm in}(n+1) C_{n+1,t-1;m} + P_{\rm out}(n-1) C_{n-1,t-1;m}
\quad (n\geq 2),
\label{e1.4a}
\end{eqnarray}
\begin{eqnarray}
C_{1,t;m} = P_{\rm in}(2) C_{2,t-1;m},
\label{e1.4b}
\end{eqnarray}
and the initial condition
\begin{eqnarray}
C_{n,0;m} = \delta_{n,m}.
\label{e1.4c}
\end{eqnarray}

The random walk described above has the advantage that the quantity $C_{n,t;m}$
is a meaningful probability for {\sl all} real values of the spatial dimension
$D$; that is, for all times $t$, the inequality
\begin{eqnarray}
0\leq C_{n,t,;m} \leq 1
\nonumber
%\label{e1.6}
\end{eqnarray}
holds. This result is in stark contrast with the random walk as it is 
conventionally defined on a hypercubic lattice \cite{r3}. For example, on a
$D$-dimensional hypercubic lattice, the probability that a random walker who is
initially at the origin ${\bf 0}$ will again be found at the origin at $t=2$ is
\begin{eqnarray}
C_{{\bf 0},2;{\bf 0}} = {1\over {2D}},
\nonumber
%\label{e1.6aa}
\end{eqnarray}
which is greater than unity for $D<{1\over 2}$; the probability that a random
walker who is initially at the origin ${\bf 0}$ will again be found at the
origin at $t=4$ is
\begin{eqnarray}
C_{{\bf 0},4;{\bf 0}} = {{6D-3}\over {8D^3}},
\nonumber
%\label{e1.6bb}
\end{eqnarray}
which is negative for $D<{1\over 2}$. 

In this paper we present an array of new results concerning $D$-dimensional
random walks. Specifically, we consider random walks that are defined by the
partial difference equations (\ref{e1.4a}), (\ref{e1.4b}) and (\ref{e1.4c}). We
show in Sec.\ \ref{s2} that there is a natural one-to-one correspondence between
the probabilities $C_{n,t;m}$ that describe a random walk on a lattice
consisting of $regions~n$, $n=1,~2,~3,~\ldots$, and a set of orthogonal
polynomials $\{ Q_{n-1}(x)\}$, $n=1,~2,~3,~\ldots$. This set of polynomials is
uniquely determined by the functions $P_{\rm in}(n)$ and $P_{\rm out}(n)$ in
Eqs. (\ref{e1.4a}), (\ref{e1.4b}) and (\ref{e1.4c}). There is a simple
expression for $C_{n,t;m}$ in terms of these polynomials. In general, one can
view a random walk on the $regions~n$ as a random sequence of raising and
lowering operators applied to the set of polynomials $\{ Q_{n-1}(x)\}$.
(Although not discussed in this paper, this correspondence between polynomials
and random walks extends to multidimensional random walks and multi-index
systems of orthonormal functions.)

If we take evenly-spaced concentric spheres ($R_n=n$), we find that for the
special cases $D=0,~1,~2$, the polynomials $\{ Q_{n-1}(x)\}$ associated with
$P_{\rm in}(n)$ and $P_{\rm out}(n)$ in Eqs. (\ref{e1.2a}), (\ref{e1.2b}) and
(\ref{e1.2c}) are standard \cite{r7,r8} classical polynomials: Gegenbauer
polynomials for $D=0$, Chebyshev polynomials for $D=1$, and Legendre polynomials
for $D=2$. However, for all other values of $D$ the polynomials have not been
previously studied and are not found in any of the usual treatments of
orthogonal polynomials. While we can generate these polynomials, we have not
been able to determine their general mathematical properties, such as their
weight function and interval of orthogonality.

In Sec.\ \ref{s3} we modify the form of $P_{\rm in}(n)$ and $P_{\rm out}(n)$ in
Eqs. (\ref{e1.2a}), (\ref{e1.2b}) and (\ref{e1.2c}) by replacing these functions
with their large-$n$ asymptotic behaviors. The polynomials that we now obtain
are well-known classical polynomials (ultraspherical polynomials) {\sl for all}
$D$. This allows us to find closed-form expressions for the probabilities
$C_{n,t;m}$ for all values of $D$.

Taking the probabilities in Sec.\ \ref{s3} we then calculate in Sec.\ \ref{s4}
extraordinarily simple, closed-form, analytic expressions for the probability of
a random walker eventually returning to the region from which the walker
started, the expected time for the walker to return to the initial region, and
other space and time moments of the probability distribution $C_{n,t;m}$. (In
contrast, in Ref.\ \cite{r2} after heavy analysis we were only able to obtain
asymptotic approximations for these moments.) We also find that for integer $D$
these moments exhibit the qualitative features ({\sl e.g.} Polya's theorem) of
random walks on $D$-dimensional hypercubic lattices.

Our long-range objective in studying $D$-dimensional random walks is to
understand critical behavior in quantum field theory. We would like to
understand, for example, the transition that occurs when a self-interacting
scalar $\phi^4$ quantum field theory in space-time dimension $D<4$ becomes a
free quantum field theory for $D>4$. One possible approach to such a problem
would be to formulate a quantum field theory in terms of random walks
\cite{r3,r4}. However, if we do so on a hypercubic lattice, it is not possible
to study these random walks except for integer values of $D$, as we have
discussed above.  As a result, we cannot use a hypercubic lattice to examine the
behavior of a quantum field theory near $D=4$. Thus, we are motivated to
investigate alternative kinds of random walks that may be consistently defined
for {\sl all real} $D$. Critical behavior has already been observed in a
two-dimensional spherically symmetric random-walk model \cite{r5}. In the next
paper in this series \cite{r6} we study this critical behavior as a continuous
function of $D$ for all $D>0$. We show in this paper that polymers adhering to
$D$-dimensional curved surfaces exhibit critical behavior.

Of course, a quantum field theory that is developed from a spherically symmetric
random walk will itself be spherically symmetric. Such a theory is physically
unacceptable because it violates causality. Nevertheless, the critical behavior
that is observed in such a theory may well be a universal function of $D$ and,
at the very least, such a theory may provide some clues as to how a scalar
quantum field theory can go from interacting to noninteracting at $D=4$. In
Sec.\ \ref{s5} we carry out some preliminary investigations of spherically
symmetric quantum field theory. Specifically, we use the machinery of
spherically symmetric random walks that is developed in Secs.\ \ref{s2} --
\ref{s4} to obtain the free two-point Green's function of a rotationally
symmetric scalar quantum theory. 

\section{CONNECTION BETWEEN POLYNOMIALS AND RANDOM WALKS}
\label{s2}
In this section we propose and discuss the following quadrature solution to
partial difference equations of the type (\ref{e1.4a}), (\ref{e1.4b}) and
(\ref{e1.4c}):
\begin{eqnarray}
C_{n,t;m}=v_{n-1}\int_{-1}^1 dx\, w(x)x^t Q_{n-1}(x)Q_{m-1}(x),
\label{e2.1}
\end{eqnarray}
where $\{ Q_n(x) \}$, $n=0,~1,~2,~\ldots~$, is a set of polynomials orthogonal 
with respect to $w(x)$ on the interval $-1\leq x\leq 1$ and $\{ v_n\}$,
$n=0,~1,~2,~\ldots~$, is a sequence of positive numbers.

The form of (\ref{e2.1}) incorporates the initial condition (\ref{e1.4c}) in a
natural way. We simply choose to {\sl normalize} the set of polynomials
$\{ Q_n(x)\}$ so that
\begin{eqnarray}
\int_{-1}^1 dx\, w(x) Q_n(x)Q_m(x)={1\over {v_n}}\delta_{n,m}.
\label{e2.2}
\end{eqnarray}
With this choice of normalization we see that at $t=0$ (\ref{e1.4c}) follows
immediately from the above statement of orthogonality:
\begin{eqnarray}
C_{n,0;m}=v_{n-1}\int_{-1}^1 dx\, w(x) Q_{n-1}(x)Q_{m-1}(x)=\delta_{n,m}.
\nonumber
%\label{e2.3}
\end{eqnarray}

We now demand that the set of polynomials $\{ Q_n(x)\}$ obey the recursion
relation
\begin{eqnarray}
P_{\rm in}(n+1) v_n Q_n(x)=v_{n-1}xQ_{n-1}(x)-P_{\rm out}(n-1)v_{n-2}Q_{n-2}
(x)\quad (n\geq 2),
\label{e2.4}
\end{eqnarray}
and the initial conditions
\begin{eqnarray}
Q_0 (x) = 1
\label{e2.5a}
\end{eqnarray}
and
\begin{eqnarray}
Q_1(x)={{v_0}\over{P_{\rm in}(2) v_1}} x.
\label{e2.5b}
\end{eqnarray}
The partial difference equations (\ref{e1.4a}), (\ref{e1.4b}) and (\ref{e1.4c})
for the probabilities $C_{n,t;m}$ are automatically satisfied so long as Eqs.\
(\ref{e2.4}), (\ref{e2.5a}) and (\ref{e2.5b}) hold.

We will now show that
\begin{eqnarray}
Q_n(1)=1
\label{e2.6}
\end{eqnarray}
for all $n\geq 0$. This interesting property is a consequence of the
conservation of probability; namely, the probability of finding the random
walker {\sl somewhere} on the lattice at an arbitrary time $t$ is unity:
\begin{eqnarray}
\sum_{n=1}^{\infty} C_{n,t;m} = 1.
\label{e2.7}
\end{eqnarray}
To establish Eq.\ (\ref{e2.7}) we merely sum Eqs.\ (\ref{e1.4a}), (\ref{e1.4b})
and (\ref{e1.4c}) over all $n\geq 1$, using Eqs.\ (\ref{e1.2c}), (\ref{e1.3}),
and (\ref{e1.4c}). Assuming that the sum
\begin{eqnarray}
f(x)\equiv \sum_{n=0}^{\infty} v_n Q_n(x)
\nonumber
%\label{e2.8}
\end{eqnarray}
exists in the space of distributions, we substitute the expression for
$C_{n,t;m}$ in Eq.\ (\ref{e2.1}) into Eq.\ (\ref{e2.7}) to obtain
\begin{eqnarray}
\int_{-1}^1 dx\, w(x) x^t Q_{m-1}(x) f(x) = 1.
\label{e2.9}
\end{eqnarray}
Next, we compute the function $f(x)$ directly from the recursion relation
(\ref{e2.4}) by summing over all $n\geq 2$, using Eqs.\ (\ref{e2.5a}) and
(\ref{e2.5b}). We obtain the following equation for $f(x)$:
\begin{eqnarray}
(1-x) f(x) = 0.
\nonumber
%\label{e1.6oo}
\end{eqnarray}
The solution to this equation is a generalized function:
\begin{eqnarray}
f(x) = \alpha \delta(x-1),
\label{e2.10}
\end{eqnarray}
where $\delta(s)$ is the Dirac delta function and $\alpha$ is a constant.
Substituting Eq.\ (\ref{e2.10}) into Eq.\ (\ref{e2.9}) gives the condition that
$Q_{m-1}(1)$ is a constant independent of $m$ for all $m\geq 1$. Finally, from
the conditions (\ref{e2.5a}) and (\ref{e2.5b}), we conclude that this constant
is 1,
\begin{eqnarray}
Q_m(1)=1,
\label{e2.11}
\end{eqnarray}
and we therefore obtain Eq.\ (\ref{e2.6}).

The result (\ref{e2.6}) enables us to find a simple formula for the set of
numbers $\{ v_n\}$. We let $x=1$ in the recursion relation (\ref{e2.4}) to
obtain
\begin{eqnarray}
P_{\rm in}(n+1)v_n=v_{n-1}-P_{\rm out}(n-1)v_{n-2}\quad (n\geq 2)
\label{e2.12}
\end{eqnarray}
and in the initial condition (\ref{e2.5b}) to obtain
\begin{eqnarray}
v_1 = {{v_0}\over{P_{\rm in}(2)}}.
\label{e2.13}
\end{eqnarray}
The unique solution to Eq.\ (\ref{e2.12}) that satisfies Eq.\ (\ref{e2.13}) is
\begin{eqnarray}
v_n=v_0\prod_{k=1}^n{{P_{\rm out}(k)}\over {P_{\rm in}(k+1)}}\quad (n\geq 1).
\label{e2.14}
\end{eqnarray}

The value of $v_0$ is determined from the orthogonality condition (\ref{e2.2})
at $n=m=0$ and the initial condition (\ref{e2.5a}):
\begin{eqnarray}
v_0 = {1\over {\int_{-1}^1 dx\, w(x)}}.
\nonumber
%\label{e2.15}
\end{eqnarray}

The result in Eq.\ (\ref{e2.14}) can be used to eliminate the numbers $v_n$ from
the recursion relation (\ref{e2.4}) giving a much simpler recursion relation for
the polynomials $Q_n(x)$:
\begin{eqnarray}
P_{\rm out}(n) Q_n(x)=xQ_{n-1}(x)-P_{\rm in}(n)Q_{n-2}(x)\quad (n\geq 2).
\label{e2.16}
\end{eqnarray}
The initial conditions in Eqs.\ (\ref{e2.5a}) and (\ref{e2.5b}) also become much
simpler:
\begin{eqnarray}
Q_0(x)=1,\quad Q_1(x)=x.
\nonumber
%\label{e2.17}
\end{eqnarray}

This recursion relation generates polynomials that exhibit parity symmetry; that
is, even-index polynomials are even functions and odd-index polynomials are odd
functions: $Q_n(-x)=(-1)^n Q_n(x)$. From the
orthogonality condition (\ref{e2.2}) one can then deduce that the weight
function $w(x)$ is an even function of $x$. As a consequence, we see from the
integral representation in Eq.\ (\ref{e2.1}) that an $n$ versus $t$ table of
values of the probabilities $C_{n,t;m}$ has a checkerboard pattern with nonzero
entries alternating with zero entries in both the $n$ and $t$ directions.
Evidently, a random walker starting from the site $m$ at $t=0$ can only reach a
site $n$ at time $t$ if $n+m+t$ is even. This parity condition is a consequence
of the original definition of our random walk in which the walker must move in
or out on every step and may not remain in the same region.

It is interesting to examine some special cases of the polynomial solution for
$C_{n,t,;m}$ in Eq.\ (\ref{e2.1}). We consider the case of equally spaced
spherical shells $R_n=n$ and look at some particular values of the dimension $D$
with $P_{\rm out}(n)$ and $P_{\rm in}(n)$ given in Eqs.\ (\ref{e1.2a}),
(\ref{e1.2b}) and (\ref{e1.2c}).

\subsection{Special case $R_n=n$, $D=1$}
\label{ss2.2}
Here,
\begin{eqnarray}
P_{\rm out}(n)={1\over 2}~~{\rm and}~~P_{\rm in}(n)={1\over 2}\quad (n\geq 2)
\label{e2.20a}
\end{eqnarray}
and
\begin{eqnarray}
P_{\rm out}(1)=1~~{\rm and}~~P_{\rm in}(1)=0.
\label{e2.20b}
\end{eqnarray}
The polynomials $\{ Q_n(x)\}$ are the standard Chebyshev polynomials of the
first kind \cite{r7}:
\begin{eqnarray}
{\cal T}_0(x)=1,\quad{\cal T}_1(x)=x,\quad {\cal T}_2(x)=2x^2-1,
\nonumber
%\label{e1.6ee}
\end{eqnarray}
\begin{eqnarray}
{\cal T}_3(x)=4x^3-3x,\quad {\cal T}_4(x)=8x^4-8x^2+1,
\nonumber
%\label{e1.6ff}
\end{eqnarray}
and so on. For these polynomials, $w(x)=1/\sqrt{1-x^2}$ and $v_n=2/\pi$,
$n\geq 1$, and $v_0=1/\pi$. The random walk probabilities in Eq.\ (\ref{e2.1})
are given by
\begin{eqnarray}
C_{n,t;m}={2\over \pi}\int_{-1}^1 dx\, {1\over\sqrt{1-x^2}}x^t {\cal T}_{n-1}
(x){\cal T}_{m-1}(x)\quad (n\geq 2),
\nonumber
%\label{e2.21a}
\end{eqnarray}
\begin{eqnarray}
C_{1,t;m}={1\over \pi}\int_{-1}^1 dx\, {1\over\sqrt{1-x^2}}x^t {\cal T}_{m-1}
(x),
\nonumber
%\label{e2.21b}
\end{eqnarray}
which for $m=1$ reduces to the particular solution
\begin{eqnarray}
C_{n,n+2j-1;1}={(n+2j-1)!\over j!(n+j-1)!2^{n+2j-2}}\quad (n\geq 2),
\nonumber
%\label{e1.6gg}
\end{eqnarray}
\begin{eqnarray}
C_{1,2t;1}={(2t)!\over t!t!2^{2t}},
\nonumber
%\label{e1.6hh}
\end{eqnarray}
given in Ref.\ \cite{r1}.

\subsection{Special case $R_n=n$, $D=2$}
\label{ss2.3}
Here,
\begin{eqnarray}
P_{\rm out}(n)={n\over{2n-1}}~~{\rm and}~~P_{\rm in}(n)={{n-1}\over{2n-1}}.
\label{e2.22}
\end{eqnarray}
The polynomials $\{ Q_n(x)\}$ are the standard Legendre polynomials \cite{r7}:
\begin{eqnarray}
{\cal P}_0(x)=1,\quad{\cal P}_1(x)=x,\quad {\cal P}_2(x)={1\over 2}(3x^2-1),
\nonumber
%\label{e1.6ii}
\end{eqnarray}
\begin{eqnarray}
{\cal P}_3(x)={1\over 2}(5x^3-3x),\quad {\cal P}_4(x)={1\over 8}(35x^4
-30x^2+3),
\nonumber
%\label{e1.6jj}
\end{eqnarray}
and so on. For these polynomials, $w(x)=1$ and $v_n = 2n+1$. Thus, the
random walk probabilities in Eq.\ (\ref{e2.1}) are given by
\begin{eqnarray}
C_{n,t;m}=(2n-1)\int_{-1}^1 dx\, x^t {\cal P}_{n-1}(x){\cal P}_{m-1}(x),
\nonumber
%\label{e2.23}
\end{eqnarray}
which for $m=1$ reduces to the particular solution
\begin{eqnarray}
C_{n,n+2j-1;1}={{(2n-1)(n+2j-1)! }\over{j!(2n+2j-1)!!2^j}},
\nonumber
%\label{e1.6kk}
\end{eqnarray}
given in Ref.\ \cite{r1}.

\subsection{Special case $R_n=n$, $D=0$}
\label{ss2.1}
Here,
\begin{eqnarray}
P_{\rm out}(n)={n-1\over 2n-1}~~{\rm and}~~P_{\rm in}(n)={n\over 2n-1}\quad
(n\geq 2)
\nonumber
%\label{e2.18a}
\end{eqnarray}
and
\begin{eqnarray}
P_{\rm out}(1)=1~~{\rm and}~~P_{\rm in}(1)=0.
\nonumber
%\label{e2.18b}
\end{eqnarray}
The first few polynomials in the set $\{ Q_n(x)\}$ are
\begin{eqnarray}
Q_0(x)=1,\quad Q_1(x)=x,\quad Q_2(x)=3x^2-2,
\nonumber
%\label{e1.6cc}
\end{eqnarray}
\begin{eqnarray}
Q_3(x)={1\over 2}(15x^3-13x),\quad Q_4(x)={1\over 6}(105x^4-115x^2+16),
\label{e1.6dd}
\end{eqnarray}
and so on. For these polynomials we have chosen $v_n=3(2n+1)/[4n(n+1)]$,
$n\geq 1$, and $v_0=3/4$. The random walk probabilities in Eq.\ (\ref{e2.1}) are
then given by
\begin{eqnarray}
C_{n,t;m}={3(2n-1)\over 4n(n-1)}\int_{-1}^1 dx\, w(x) x^t Q_{n-1} (x) Q_{m-1}(x)
\quad (n\geq 2),
\nonumber
%\label{e2.19a}
\end{eqnarray}
\begin{eqnarray}
C_{1,t;m}={3\over 4}\int_{-1}^1 dx\, w(x) x^t Q_{m-1}(x).
\nonumber
%\label{e2.19b}
\end{eqnarray}
The polynomials Eq.\ (\ref{e1.6dd}) are closely related to the standard
Gegenbauer (ultraspherical) polynomials, $\{ {\cal C}_n^{(\alpha})(x)\}$, with
upper index $\alpha = 3/2$ \cite{r7}. These particular Gegenbauer polynomials
satisfy the recursion relation
\begin{eqnarray}
(n+1){\cal C}_{n+1}^{(3/2)}(x)=(2n+3)x {\cal C}_n^{(3/2)}(x)-(n+2)
{\cal C}_{n-1}^{(3/2)}(x)\quad (n > 0)
\nonumber
%\label{e2.19aa}
\end{eqnarray}
and the initial conditions ${\cal C}_0^{(3/2)}(x)=1$ and ${\cal C}_1^{(3/2)}(x)
=3x$. These Gegenbauer polynomials are orthogonal on the interval $-1\leq
x\leq 1$ with respect to the weight function $w(x) = 1-x^2$. The polynomial
$Q_{n+1}(x)$ satisfies the same recursion relation as these Gegenbauer
polynomial ${\cal C}_n^{(3/2)}(x)$. However, it is generated from different
initial conditions. We have been able to show that the weight function $w(x)$
with respect to which the set of polynomials $\{ Q_n(x) \}$ is orthogonal
satisfies the integral equation
\begin{eqnarray}
\int_{-1}^1 dt\, {w(t)\over 1-xt^2} = {2\sqrt{x} \over (1-x)[ 
\ln (1+\sqrt{x})- \ln (1-\sqrt{x})]}.
\nonumber
%\label{e2.19ii}
\end{eqnarray}
We do not know a closed-form solution to this equation.

\subsection{Special case $R_n=n$, $D=3$}
\label{ss2.4}
Now,
\begin{eqnarray}
P_{\rm out}(n)={{n^2}\over{2n^2-2n+1}}~~{\rm and}~~P_{\rm in}(n)={{(n-1)^2}
\over{2n^2-2n+1}}.
\nonumber
%\label{e2.24}
\end{eqnarray}
For this case we can calculate any finite number of polynomials $\{ Q_n(x)\}$:
\begin{eqnarray}
Q_0(x)=1,\quad Q_1(x)=x,\quad Q_2(x)={1\over 4}(5x^2-1),
\nonumber
%\label{e1.6ll}
\end{eqnarray}
\begin{eqnarray}
Q_3(x)={1\over 36}(65x^3-29x),\quad Q_4(x)={1\over 576}(1625x^4-1130x^2+81),
\nonumber
%\label{e1.6mm}
\end{eqnarray}
and so on. These polynomials {\sl are} orthogonal and they satisfy the
normalization constraint (\ref{e2.11}). However, for this value of $D$ (and for
all values of $D$ other than $D=0,~1,~2$) these polynomials are not related
to the standard classical polynomials that one can find in reference books. We
are unable to determine analytically the weight function $w(x)$ with respect to
which these polynomials are orthogonal! Thus, the formal expression
Eq.\ (\ref{e2.1}) for the probabilities $C_{n,t;m}$ is not very useful.

In the next section we devise a random walk process for which we {\sl can}
determine the weight function and thus find in closed form physically realistic
probabilities $C_{n,t;m}$ for all values of $D>0$.

\section{RANDOM WALKS FOR ULTRASPHERICAL POLYNOMIALS}
\label{s3}
In this section we show how to modify the expressions for $P_{\rm out}(n)$ and
$P_{\rm in}(n)$ in Eqs.\ (\ref{e1.2a}), (\ref{e1.2b}) and (\ref{e1.2c}) so that
we are able to obtain analytic closed-form expressions for $C_{n,t;m}$ for
{\sl all} values of $D>0$ for the case of evenly-spaced spherical shells
$R_n =n$. The random walk process examined in Subsection \ref{ss2.4} is too
difficult to solve in closed form simply because the formulas for
$P_{\rm out}(n)$ and $P_{\rm in}(n)$ in Eqs.\ (\ref{e1.2a}), (\ref{e1.2b}) and
(\ref{e1.2c}) become much too complicated when $D$ takes on values other than 0,
1, or 2.

As we will see, the polynomials generated by the recursion relation
(\ref{e2.16}) belong to a set of well-known classical polynomials if we take the
formulas for $P_{\rm out}(n)$ and $P_{\rm in}(n)$ to be bilinear functions of
$n$ of the general form
\begin{eqnarray}
P_{\rm out}(n)={an+b\over cn+d},
\label{e3.1}
\end{eqnarray}
with $P_{\rm in}(n)=1-P_{\rm out}(n)$. Note that bilinear functions contain
{\sl three} arbitrary parameters. We fix these parameters as follows. First, we
demand that the random walk be confined to the values of $n\geq 1$. To impose
this condition we require that $P_{\rm out}(1)=1$, or, equivalently that
$P_{\rm in} (1)=0$. This fixes one parameter. Second, we demand that the
large-$n$ asymptotic behavior of $P_{\rm out}(n)$ in Eqs.\ (\ref{e1.2a}) and
(\ref{e3.1}) agree to order $n$. These two conditions above yield the unique
choice
\begin{eqnarray}
P_{\rm out}(n)={{n+D-2}\over{2n+D-3}}~~{\rm and}~~P_{\rm in}(n)={{n-1}
\over{2n+D-3}}.
\label{e3.2}
\end{eqnarray}

By determining the arbitrary parameters in Eq.\ (\ref{e3.1}) at the two boundary
points $n=1$ and $n=\infty$ we obtain a uniformly accurate approximation to 
$P_{\rm out}(n)$ and $P_{\rm in}(n)$ in Eqs.\ (\ref{e1.2a}), (\ref{e1.2b}) and
(\ref{e1.2c}) for all $n\geq 1$. In fact, Eq.\ (\ref{e3.2}) agrees exactly with
Eqs.\ (\ref{e1.2a}), (\ref{e1.2b}) and (\ref{e1.2c}) for $D=1$ [see Eqs.\
(\ref{e2.20a}) and (\ref{e2.20b})] and $D=2$ [see Eq.\ (\ref{e2.22})]. For other
values of $D$, Eq.\ (\ref{e3.2}) continues to be a good approximation, as is
verified in Fig.\ \ref{f1}, where we compare $P_{\rm in}(n)$ in Eq.\ (\ref
{e1.2b}) with $P_{\rm in}(n)$ in Eq.\ (\ref{e3.2}) for several values of $D$.

The requirement that Eqs.\ (\ref{e1.2a}) and (\ref{e3.1}) agree to order $n$ as
$n\to\infty$ incorporates the crucial dependence upon the dimension $D$ of
space; to wit, as $D$ increases, a random walker is more likely to move outward
than to move inward. As we will see in Sec.\ \ref{s4}, it is this bias that
gives rise to Polya's theorem; this theorem states that for $0<D\leq 2$ a random
walker returns to the starting point with probability 1, while for $D>2$ this
probability is less than 1.

Substituting the formulas above into Eq.\ (\ref{e2.16}) gives the recursion
relation
\begin{eqnarray}
(n+D-2)Q_n(x)=(2n+D-3)xQ_{n-1}(x)-(n-1)Q_{n-2}(x)\quad (n\geq 2).
\label{e3.3}
\end{eqnarray}
Taking as initial conditions
\begin{eqnarray}
Q_0(x)=1,
\nonumber
%\label{e3.4a}
\end{eqnarray}
\begin{eqnarray}
Q_1(x)=x,
\nonumber
%\label{e3.4b}
\end{eqnarray}
we can easily use Eq.\ (\ref{e3.3}) to generate subsequent polynomials:
\begin{eqnarray}
Q_2(x)={1\over D}[(D+1)x^2-1],
\nonumber
%\label{e3.4c}
\end{eqnarray}
\begin{eqnarray}
Q_3(x)={1\over D}[(D+3)x^3-3x],
\nonumber
%\label{e3.4d}
\end{eqnarray}
\begin{eqnarray}
Q_4(x)={1\over D^2+2D}[(D^2+8D+15)x^4-(6D+18)x^2+3],
\nonumber
%\label{e3.4e}
\end{eqnarray}
\begin{eqnarray}
Q_5(x)={1\over D^2+2D}[(D^2+12D+35)x^5-(10D+50)x^3+15x].
\nonumber
%\label{e3.4f}
\end{eqnarray}
These polynomials are just the Gegenbauer (ultraspherical) polynomials \cite{r7}
${\cal C}_n^{(\alpha)}(x)$ normalized so that $Q_n(1)=1$:
\begin{eqnarray}
Q_n(x)={n!\Gamma(D-1)\over\Gamma(n+D-1)}{\cal C}_n^{ (
{D-1\over 2} )}(x).
\label{e3.5}
\end{eqnarray}
Gegenbauer polynomials are hypergeometric functions in the variable $(x-1)/2$;
furthermore, since they are polynomials, the Taylor series for $Q_n(x)$ about
$x=1$ terminates:
\begin{eqnarray}
Q_n(x)=\sum_{j=0}^n \left ( {n \atop j} \right ) {\Gamma(j+D+n-1)\Gamma(D/2)
\over\Gamma(D+n-1)\Gamma(j+D/2)} \left ( {x-1\over 2}\right ) ^j.
\label{e3.5.a}
\end{eqnarray}

 From the conventional theory of Gegenbauer polynomials \cite{r7} we immediately
know the weight function with respect to which the polynomials $Q_n(x)$ are
orthogonal:
\begin{eqnarray}
w(x)= (1-x^2)^{(D-2)/2}.
\label{e3.6}
\end{eqnarray}
Also, the normalization coefficients $v_n$ in Eq.\ (\ref{e2.2}) are identified
as
\begin{eqnarray}
v_n={{(2n+D-1)\Gamma(n+D-1)\Gamma[(D+1)/2]}\over{\sqrt{\pi}n!\Gamma(D/2)
\Gamma(D)}}.
\label{e3.7}
\end{eqnarray}
 Finally, we note that the polynomials $Q_n(x)$ satisfy the Sturm-Liouville
eigenvalue differential equation
\begin{eqnarray}
\left [(1-x^2){{d^2}\over{dx^2}}-Dx{d\over{dx}}+n(n+D-1)\right ] Q_n(x)= 0
\nonumber
%\label{e3.8}
\end{eqnarray}
and the first-order difference-differential equation
\begin{eqnarray}
\left [ (1-x^2){{d}\over{dx}}+nx\right ] Q_n(x)=nQ_{n-1}.
\label{e3.9}
\end{eqnarray}

Now that we have identified explicitly the polynomials $Q_n(x)$, the weight
function $w(x)$, and the normalization coefficients $v_n$, we can use the
formula in Eq.\ (\ref{e2.1}) to calculate the moments of $C_{n,t;m}$ and obtain
a physical description of our random walk.

\section{QUANTITATIVE DESCRIPTION OF THE RANDOM WALK}
\label{s4}
In this section we discuss the properties of the hyperspherical random walk
introduced in Sec.\ \ref{s3}. We calculate the probability of eventually
returning to the starting point of a random walk, the expected time of return,
and various other moments of the random walk probabilities. As will be evident,
the key advantage of this random walk is that all of these quantities can be
calculated {\sl in closed form}.

\subsection{Probability of eventual return}
\label{ss4.1}
In a physical description of a random walk the simplest and most natural
question to ask is, what is the probability of eventually returning to the
starting point? The probability that a random walker will {\sl eventually}
return to $region~m$, given that the walker started in $region~m$, is denoted
$\Pi_m(D)$. To calculate $\Pi_m(D)$ we use generating function methods
previously described [see Eq.\ (2.11) of Ref.\ \cite{r1}]; to wit,
\begin{eqnarray}
\Pi_m(D)=1-{1\over\sum_{t=0}^{\infty}C_{m,2t;m}}.
\label{e4.1}
\end{eqnarray}

Our problem is now to evaluate the sum, which we denote by $S_m$, in Eq.\
(\ref{e4.1}). Using Eq.\ (\ref{e2.1}) with $Q_n(x)$ given in Eq.\ (\ref{e3.5})
we have
\begin{eqnarray}
S_m=\sum_{t=0}^{\infty}C_{m,2t;m}=\sum_{t=0}^{\infty}v_{m-1}\int_{-1}^{1}dx\;
(1-x^2)^{(D-2)/2} x^{2t} \left [Q_{m-1}(x)\right ]^2.
\label{e4.2}
\end{eqnarray}

Note that this sum is divergent unless $D>2$. To verify this assertion we
observe that the large-$t$ asymptotic behavior of the integral in Eq.\
(\ref{e4.2}) is
\begin{eqnarray}
\int_{-1}^{1}dx\; (1-x^2)^{(D-2)/2} x^{2t} \left [Q_{m-1}(x)\right ]^2
\sim\Gamma (D/2)t^{-D/2}\quad (t\to\infty).
\nonumber
%\label{e1.6nn}
\end{eqnarray}
Thus,
\begin{eqnarray}
\Pi_m(D)=1 \quad (0<D\leq 2).
\label{e4.3}
\end{eqnarray}

When $D>2$ the sum $S_m$ converges, and we begin the evaluation by
interchanging the order of summation and integration:
\begin{eqnarray}
S_m=v_{m-1}\int_{-1}^{1}dx\; (1-x^2)^{(D-4)/2}\left [Q_{m-1}(x)\right ]^2.
\nonumber
%\label{e4.4}
\end{eqnarray}
We evaluate this integral exactly using the recursion relation (\ref{e3.3}) and
the difference-differential equation (\ref{e3.9}). The result is
\begin{eqnarray}
S_m={2m+D-3\over D-2}.
\label{e4.5}
\end{eqnarray}
Substituting into Eq.\ (\ref{e4.1}) gives
\begin{eqnarray}
\Pi_m(D)={2m-1\over 2m+D-3}\quad (D>2).
\label{e4.6}
\end{eqnarray}
This result is {\sl exact} for all $m$ and $D$ \cite{wow}. The probability in
Eqs.\ (\ref{e4.3}) and (\ref{e4.6}) confirms that Polya's theorem holds for this
model of a random walk, regardless of the region in which the walk begins.

 For a hypercubic lattice the probability of eventually returning to the
starting point of a random walk is given in terms of an integral [see
Eq.\ (2.12) of Ref.\ \cite{r1}]:
\begin{eqnarray}
\Pi(D)=1-{1\over\int_{0}^{\infty}dt\; e^{-t}[I_0(t/D)]^D},
\nonumber
%\label{e4.7}
\end{eqnarray}
where $I_0(x)$ is the modified Bessel function. Unlike the random walk discussed
in this paper, when $D>2$, $\Pi(D)$ cannot be given in closed form (except for
the special case $D=3$). However, the asymptotic expansion of $\Pi(D)$ for large
$D$ is known [see Eq.\ (2.15) of Ref.\ \cite{r1}]:
\begin{eqnarray}
\Pi(D)\sim {1\over 2D}+{1\over 2D^2}+\ldots\quad (D\to\infty).
\nonumber
%\label{e4.8}
\end{eqnarray}
Note that for large $D$, the probability of returning to the starting point
of a random walk falls off algebraically like $1/D$ in both models. In contrast,
for the hyperspherical random walk discussed in Ref.\ \cite{r1}, the probability
function $\Pi_1(D)=1-1/\zeta(D-1)$ falls off {\sl exponentially} like $2^{1-D}$
for large $D$. As functions of $D$, the hypercubic $\Pi(D)$ and the
hyperspherical $\Pi_m(D)$ discussed here both exhibit cusps at $D=2$.

Observe that for large $m$, $\Pi_m(D)$ approaches $1$. This happens because the
available entropy for the random walk becomes constant; at large radius a sphere
looks locally like a plane. Indeed, as $n\to\infty$, the recursion relation
(\ref{e3.3}) approaches that of a one-dimensional random walk for which
$P_{\rm out}(n)=P_{\rm in}(n)={1\over 2}$.

\subsection{Expected time of return}
\label{ss4.2}
As explained in Ref.\ \cite{r2}, the expected time of return $T_m(D)$ of a
random walker who begins the walk in $region~m$ is obtained from the first
moment of $C_{m,t;m}$:
\begin{eqnarray}
T_m(D)={\sum_{t=0}^{\infty} 2tC_{m,2t;m}\over \Pi_m(D)(\sum_{t=0}^{\infty}
C_{m,2t;m})^2}.
\label{e4.9}
\end{eqnarray}
Again, using formulas (\ref{e3.3}) and (\ref{e3.9}) we can calculate the sums
in Eq.\ (\ref{e4.9}) straightforwardly. We find that
\begin{eqnarray}
T_m(D)=\cases{
\infty& $(0<D\leq 4)$,\cr\cr
{2(D-2)[(2m-1)D+2(m-1)(m-2)]\over (2m-1)D(D-4)}& $(D>4)$.\cr}
\nonumber
%\label{e4.10}
\end{eqnarray}

Note that as $D$ increases, $T_m(D)$ approaches 2, independent of the starting
point $m$. This is because for very large dimension $D$, if a random walker does
not return to the starting point on the second step, the random walker will
{\sl never} return; as $D\to\infty$ the entropy for moving outward dominates the
walk. However, for fixed $D$ as $m$ increases $T_m(D)$ diverges. This is because
for large $m$ the $D$-dimensional walk approaches a one-dimensional walk for
which the expected time of return is infinite.

\subsection{Higher Temporal Moments}
\label{ss4.3}
In general, all temporal moments can be calculated in closed form. The $p$th
temporal moment $\sum_{t=0}^{\infty} t^p C_{m,2t;m}$ is a rational function of
$D$ and $m$ whose complexity increases with $p$. The sum defining the $p$th
temporal moment converges when $D>2p+2$ and diverges when $D\leq 2p+2$. We
list the first four temporal moments below. [Note that the zeroth moment,
$S_m$, is already given in Eq.\ (\ref{e4.5}).]
\begin{eqnarray}
\sum_{t=0}^{\infty} C_{m,2t;m}=S_m={(2m+D-3) M_0 \over D-2},
\nonumber
%\label{e4.3.1}
\end{eqnarray}
\begin{eqnarray}
\sum_{t=0}^{\infty} t C_{m,2t;m}={(2m+D-3) M_1 \over (D-4) (D-2) D},
\nonumber
%\label{e4.3.2}
\end{eqnarray}
\begin{eqnarray}
\sum_{t=0}^{\infty} t^2 C_{m,2t;m}={(2m+D-3) M_2 \over (D-6)(D-4)(D-2)D(D+2)},
\nonumber
%\label{e4.3.3}
\end{eqnarray}
\begin{eqnarray}
\sum_{t=0}^{\infty} t^3 C_{m,2t;m}={(2m+D-3) M_3 \over
(D-8) (D-6) (D-4) (D-2) D (D+2) (D+4)},
\nonumber
%\label{e4.3.4}
\end{eqnarray}
where
\begin{eqnarray}
M_0&=&1,\nonumber\\
M_1&=&(2m-1)D+2(m-1)(m-2),\nonumber\\
M_2&=&(2m-1)D^3+2(7m^2-13m+7)D^2+4(m-1)(6m^2-17m+16)D\nonumber\\
&&\qquad +12(m-1)(m-2)(m^2-3m+4),\nonumber\\
M_3&=&(2m-1)D^5+2(19m^2-29m+15)D^4+2(96m^3-296m^2+386m-173)D^3\nonumber\\
&&\qquad +4(99m^4-486m^3+1093m^2-1184m+477)D^2\nonumber\\
&&\qquad +4(90m^5-621m^4+2040m^3-3683m^2+3414m-1252)D\nonumber\\
&&\qquad +24(m-1)(m-2)(5m^4-30m^3+97m^2-156m+104).\nonumber
%\label{e4.3.5}
\end{eqnarray}

\subsection{Spatial Moments}
\label{ss4.4}
The $k$th spatial moment of a random walk is defined as a weighted average
over the probabilities $C_{n,t;m}$:
\begin{eqnarray}
\langle R^k\rangle _t \equiv \sum_{n=1}^{\infty} n^k C_{n,t;m}.
\label{e4.x1}
\end{eqnarray}
Note that in general $\langle R^k\rangle _t$ depends on the starting point $m$
of the random walk. We have suppressed the argument $m$ because, as we will
see, the leading asymptotic behavior of $\langle R^k\rangle _t$ as $t\to\infty$
and the first correction to this behavior is independent of $m$. (The second
correction does depend on $m$.)

We have found an {\sl exact} expression for $\langle R^k\rangle _t$ for all
values of $t$ for the special case $m=1$:
\begin{eqnarray}
\langle R^k\rangle_{2t}=(2t+1)^k+\sum_{r=1}^t \sum_{s=r+1}^{t+1}(-1)^{r+s}
f(2r-1,2s-1,2t,k)
\label{e4.x2a}
\end{eqnarray}
and
\begin{eqnarray}
\langle R^k\rangle_{2t+1}=(2t+2)^k+\sum_{r=1}^t\sum_{s=r+1}^{t+1}(-1)^{r+s}
f(2r,2s,2t+1,k),
\label{e4.x2b}
\end{eqnarray}
where
\begin{eqnarray}
f(x,y,t,k)={(x-y)\Gamma(x+y-1)\Gamma(t+1)(x^k-y^k)\over 2^t\Gamma(x)\Gamma(y)
\Gamma\left ({x+y\over 2}\right )\Gamma\left ({t-x+3\over 2}\right )
\Gamma\left ({t-y+3\over 2}\right ) (D+x+y-3)}.
\label{e4.x2c}
\end{eqnarray}

This formula has the virtue that the $D$ dependence is very simple; the
parameter $D$ occurs just once in the denominator of $f$ in Eq.\ (\ref{e4.x2c}).
Furthermore, for the special case of the zeroth moment, setting $k=0$ in Eqs.\
(\ref{e4.x2a}) or (\ref{e4.x2b}) immediately gives the result
$\langle R^0\rangle_{t}=1$, which states that probability is conserved. For
$k>1$ this formula is inherently complicated. It is not easy to determine the
asymptotic behavior of $\langle R^k\rangle_{t}$ for large $t$ from Eqs.\
(\ref{e4.x2a}) or (\ref{e4.x2b}) because terms in the double sum oscillate in
sign.

To find the asymptotic behavior of $\langle R^k\rangle_{t}$ as $t\to\infty$ we
use generating-function techniques. We rewrite Eq.\ (\ref{e4.x1}) as a
derivative operator applied $k$ times to a power series:
\begin{eqnarray}
\langle R^k\rangle_t=\lim_{z\to 1} (z{d\over dz})^k \sum_{n=1}^{\infty} z^n
C_{n,t;m}.
\label{e4.x3}
\end{eqnarray}
Next, we substitute into Eq.\ (\ref{e4.x3}) the integral representation for the
probability $C_{n,t;m}$ in Eq.\ (\ref{e2.1}) and use Eq.\ (\ref{e3.6}). We
obtain
\begin{eqnarray}
\langle R^k\rangle_t=\lim_{z\to 1} \int_{-1}^1 dx\; (1-x^2){(D-2)/2}x^t
Q_{m-1}(x) (z{d\over dz})^k \left [ z \sum_{n=0}^{\infty} z^n v_n Q_n(x)\right].
\label{e4.x4}
\end{eqnarray}
It is convenient to use the expression for $v_n$ in Eq.\ (\ref{e3.7}) and the
recursion relation (\ref{e3.3}) for $Q_n(x)$ to evaluate the sum in
Eq.\ (\ref{e4.x4}):
\begin{eqnarray}
\sum_{n=0}^{\infty} z^n v_n Q_n(x)={\Gamma\left ({D+1\over 2}\right )\over
\Gamma\left ({1\over 2}\right )\Gamma\left ({D\over 2}\right ) }
(1-z^2) (1-2xz+z^2)^{-(D+1)/2}.
\nonumber
%\label{e4.x5}
\end{eqnarray}

We are interested in the behavior of the resulting integral as $t\to\infty$. By
Laplace's method this integral is dominated by values of $x$ near 1 in this
limit. Thus, for fixed $m$ Eq.\ (\ref{e2.11}) implies that we may replace
$Q_{m-1}(x)$ by 1 to leading order; we thus conclude that the leading asymptotic
behavior of $\langle R^k\rangle_t$ is independent of $m$. To obtain higher-order
terms in the asymptotic expansion we replace $Q_{m-1}(x)$ by the expansion in
Eq.\ (\ref{e3.5.a}). A straightforward asymptotic analysis of the resulting
integral gives the first few terms in the asymptotic expansion of
$\langle R^k\rangle_t$ for large $t$ with $m$ fixed:
\begin{eqnarray}
\langle R^k\rangle_t \sim&& {\Gamma\left ({D+k\over 2}\right )\over
\Gamma\left ({D\over 2}\right )} (2t)^{k/2}
\Bigg \{ 1-{k(D-3)\Gamma\left ( {D+k-1\over 2}\right ) \over 2 \Gamma \left (
{D+k\over 2}\right )}(2t)^{-1/2}+k \Bigg [ {(m-1)(m+D-2)\over D} \nonumber\\
&& + {3D^2k-18Dk+12D-2k^2+33k-28\over 12(D+k-2)} \Bigg ] (2t)^{-1}
+ {\cal O}\left ( t^{-3/2}\right )\Bigg \}\quad (t\to\infty).
\label{e4.x6}
\end{eqnarray}
Observe that the leading term in this asymptotic expansion is precisely the same
as the result in Eq.\ (3.4) of Ref.\ \cite{r2} for the case of spherically
symmetric random walks described by the probabilities in Eqs.\ (\ref{e1.2a}),
(\ref{e1.2b}) and (\ref{e1.2c}) with $R_n=n$. The result in Eq.\ (\ref{e4.x6})
is obtained directly and with considerably less effort than that in Ref.\
\cite{r2}, where only the leading asymptotic behavior was obtained. Note that
the first two terms in Eq.\ \ref{e4.x6} are independent of the starting point
$m$. To verify the accuracy of this asymptotic expansion we compare the first
three partial sums of this series with the exact values of the moments obtained
numerically at $t=1000$; this comparison is given in Tables \ref{t1} and
\ref{t2}. In Table \ref{t1} we consider the $k$th moment for various values of
$k$ and $D$ with $m=1$. In Table \ref{t2} we consider the first and second
moments for various values of the starting point $m$ with $D=2$.

 From the asymptotic behavior in Eq.\ (\ref{e4.x6}) with $k=2$ we can determine
the Hausdorff dimension $D_H$ of the random walk \cite{r2}. We find that
\begin{eqnarray}
D_H = 2
\nonumber
%\label{e4.x7}
\end{eqnarray}
for all values of $D$. This result agrees with that obtained in \cite{r3} for
a $D$-dimensional hypercubic lattice.

\section{APPLICATION TO QUANTUM FIELD THEORY}
\label{s5}
One of our long-range goals in our study of $D$-dimensional random walks is a
deeper understanding of $D$-dimensional quantum field theory. In particular,
we are interested in how critical phenomena in such theories depend on the
dimension of space-time. We are especially interested in how a $\phi^4$ scalar
field theory becomes free as $D\to 4$. We have already conducted several
investigations of $D$-dimensional quantum-mechanical and field-theoretic systems
\cite{r9,r10,r11,r12}. In this section, as an elementary illustration of how to
apply our work on $D$-dimensional random walks to quantum field theory, we use
the random walk probabilities $C_{n,t;m}$ in Eq.\ (\ref{e2.1}) with the
polynomials $Q_n(x)$ given in Eq.\ (\ref{e3.5}) to calculate the Euclidean
two-point Green's function of a $D$-dimensional free scalar quantum field theory
having spherical symmetry. We will then verify our calculation by taking the
spherical average of the two-point Green's function of a conventional
translationally invariant (nonspherically symmetric) Euclidean field theory
\cite{Sharp}.

\subsection{Derivation of spherically symmetric propagator from random walk
probabilities $C_{n,t;m}$}
\label{ss5.1}

For this calculation we follow the standard recipe discussed in Ref.\ \cite{r3}.
Specifically, we begin with the generating function $G(n,m,\lambda)$ for the
temporal moments of the probabilities $C_{n,t;m}$:
\begin{eqnarray}
G(n,m,\lambda)=\sum_{t=0}^{\infty} \lambda^t C_{n,t;m}.
\nonumber
%\label{e5.1}
\end{eqnarray}
Our objective is to find the continuum limit of this expression.

For definiteness we choose $n$, $m$, and $t$ to be even: $n=2N$, $m=2M$, $t=2T$.
Also, without loss of generality, we take $N\geq M$. Substituting the formula
for $C_{m,t;m}$ in Eq.\ (\ref{e2.1}) with $Q_n(x)$ given by ultraspherical
polynomials in Eq.\ (\ref{e3.5}), and $w(x)$ in Eq.\ (\ref{e3.6}), and $v_n$ in
Eq.\ (\ref{e3.7}), we obtain
\begin{eqnarray}
G(2N,2M,\lambda)={4N\Gamma^2\left({D-1\over 2}\right)\over\pi M^{D-2}}
\int_0^1 dx\; (1-x^2)^{D-2\over 2}\sum_{T=N-M}^{\infty}(x\lambda)^{2T}
{\cal C}_{2N-1}^{({D-1\over 2})}(x) {\cal C}_{2M-1}^{({D-1\over 2})}(x).
\label{e5.2}
\end{eqnarray}

Next, we perform the sum in Eq.\ (\ref{e5.2}):
\begin{eqnarray}
G(2N,2M,\lambda)={4N\Gamma^2\left({D-1\over 2}\right)\over\pi M^{D-2}}
\int_0^1 dx\; (1-x^2)^{D-2\over 2}{(x\lambda)^{2N-2M}\over 1-x^2 \lambda^2}
{\cal C}_{2N-1}^{({D-1\over 2})}(x) {\cal C}_{2M-1}^{({D-1\over 2})}(x).
\label{e5.3}
\end{eqnarray}

To prepare for taking the continuum limit we make use of the equivalence of
Gegenbauer and Jacobi polynomials \cite{r7}:
\begin{eqnarray}
{\cal C}_{2N-1}^{({D-1\over 2})}(x)={\Gamma(D/2)\Gamma(2N-2+D)\over
\Gamma(D-1)\Gamma(2N-1+D/2)}{\cal P}_{2N-1}^{({D-2\over 2},{D-2\over 2})}(x).
\label{e5.4}
\end{eqnarray}
Substituting Eq.\ (\ref{e5.4}) into Eq.\ (\ref{e5.3}) and taking $N$ and $M$
large gives
\begin{eqnarray}
G(2N,2M,\lambda)&=&2^{4-D} N^{D/2} M^{1-D/2}
\int_0^1 dx\; (1-x^2)^{D-2\over 2}{(x\lambda)^{2N-2M}\over 1-x^2 \lambda^2}
\nonumber\\
&\times& {\cal P}_{2N-1}^{({D-2\over 2},{D-2\over 2})}(x)
{\cal P}_{2M-1}^{({D-2\over 2},{D-2\over 2})}(x).
\label{e5.5}
\end{eqnarray}

When $N$ and $M$ are large the integral in Eq.\ (\ref{e5.5}) is dominated by
values of $x$ near 1. Thus, we make the change of variable $x=1-\epsilon^2
s^2/2$, where $\epsilon$ is a small parameter:
\begin{eqnarray}
G(2N,2M,\lambda)&=&4 \epsilon^2 N^{D/2} M^{1-D/2}
\int_0^{\sqrt{2}/\epsilon} ds\;
{s[\lambda (1-\epsilon^2 s^2/2)]^{2N-2M}\over 1-\lambda^2 (1-\epsilon^2s^2/2)^2}
\nonumber\\
&\times& \left ( {\epsilon s\over 2}\right )^{D-2\over 2}
{\cal P}_{2N-1}^{({D-2\over 2},{D-2\over 2})}\left (1-{\epsilon^2 s^2\over
2} \right ) \left ( {\epsilon s\over 2}\right )^{D-2\over 2} {\cal P}_{2M-1}^{(
{D-2\over 2},{D-2\over 2})}\left (1-{\epsilon^2 s^2\over 2} \right ).
\label{e5.6}
\end{eqnarray}

We now make use of the following asymptotic limit for Jacobi polynomials
\cite{r7}:
\begin{eqnarray}
\lim_{\eta \to 0} \left ( {\eta s\over 2}\right )^{\alpha}
{\cal P}_{1/\eta}^{(\alpha,\alpha)} \left (1-{\eta^2 s^2\over 2} \right )
=J_{\alpha}(s),
\nonumber
%\label{e5.7}
\end{eqnarray}
where $J_{\alpha}(s)$ is a Bessel function. Because $N$ and $M$ are large, we
can use this asymptotic limit twice in Eq.\ (\ref{e5.6}):
\begin{eqnarray}
G(2N,2M,\lambda)&=&4 \epsilon^2 N^{D/2} M^{1-D/2}
\int_0^{\sqrt{2}/\epsilon} ds\;
{s[\lambda (1-\epsilon^2 s^2/2)]^{2N-2M}\over 1-\lambda^2 (1-\epsilon^2s^2/2)^2}
\nonumber\\
&\times& J_{D-2\over 2} [\epsilon (2N-1)s] J_{D-2\over 2} [\epsilon (2M-1)s].
\label{e5.8}
\end{eqnarray}

We introduce the continuum variables $r$ and $r'$ by
\begin{eqnarray}
\mu r = \epsilon (2N-1), \qquad \mu r'= \epsilon (2M-1),
\nonumber
%\label{e5.9}
\end{eqnarray}
where $\mu$ is a mass parameter. Note that $r>r'$. Also, since $\epsilon << 1$,
we may replace the upper limit of integration in Eq.~(\ref{e5.8}) by $\infty$
and simplify the integrand:
\begin{eqnarray}
G(2N,2M,\lambda)=2 \epsilon r^{D/2} (r')^{1-D/2}\mu
\int_0^{\infty} ds\; {s\over 1-\lambda^2 (1-\epsilon^2 s^2)}
J_{D-2\over 2} (\mu rs) J_{D-2\over 2} (\mu r' s).
\nonumber
%\label{e5.10}
\end{eqnarray}

Finally, we make use of the following Bessel function integral identity
\cite{r13}:
\begin{eqnarray}
\int_0^{\infty} ds\; {s\over s^2 + c^2} J_{\nu}(as) J_{\nu}(bs)=I_{\nu}(bc)
K_{\nu}(ac)\quad (a>b),
\nonumber
%\label{e5.11}
\end{eqnarray}
where $I_{\nu}$ and $K_{\nu}$ are modified Bessel functions. Taking,
$\lambda^2 \epsilon^2 = 1-\lambda^2$, we have
\begin{eqnarray}
G(2N,2M,\lambda)={2 \mu\over \epsilon} r^{D/2} (r')^{1-D/2}
I_{D-2\over 2} (\mu r') K_{D-2\over 2} (\mu r).
\label{e5.12}
\end{eqnarray}

Apart from a multiplicative normalization constant, the expression in
Eq.~(\ref{e5.12}) is the final result for the Euclidean propagator. Let
${\cal G}(r\to r')$ represent the spherically averaged amplitude for a free
scalar particle of mass $\mu$ to propagate from some point on a sphere of radius
$r$ to some point on a sphere of radius $r'$. Note that this probability
amplitude is not symmetric under the interchange of $r$ and $r'$; when $D>1$ it
is more likely for a particle to propagate from a sphere of smaller radius to a
sphere of larger radius than for the reverse to occur. This is because the final
state of the particle propagating to the larger sphere has a higher entropy.
This asymmetry does not occur in translationally invariant theories. Our final,
properly normalized, result for the propagator is
\begin{eqnarray}
{\cal G}(r\to r')=(r')^{D-1} (rr')^{1-D/2}
I_{D-2\over 2} (\mu r_<) K_{D-2\over 2} (\mu r_>),
\label{e5.13}
\end{eqnarray}
where 
\begin{eqnarray}
r_> = {\rm max}\{ r,r'\}\quad {\rm and}\quad r_< = {\rm min}\{ r,r'\}.
\nonumber
%\label{e5.14}
\end{eqnarray}
The normalization of the Green's function in Eq.~(\ref{e5.13}) will be verified
in Sec.~\ref{ss5.2}.

The propagation asymmetry in Eq.~(\ref{e5.13}) is a continuum manifestation of
the directional bias that is present in spherically symmetric random walks. Note
that $C_{n,t;m}$, the probability of walking from $m$ to $n$ [see
Eq.~(\ref{e2.1})], is {\sl not} a symmetric function of $m$ and $n$. Rather, it
is the function $v_{n-1}$ in Eq.~(\ref{e3.7}) multiplying a symmetric function
of $m$ and $n$. The function $v_{n-1}$ represents the random-walk entropy
associated with the volume of hyperspherical {\sl region} $n$. The asymmetry in
Eq.~(\ref{e5.13}) is a direct consequence of the asymmetry in $C_{n,t;m}$.

\subsection{Normalization of the two-point Green's function}
\label{ss5.2}

The free propagator in momentum space for a $D$-dimensional translationally
symmetric scalar field theory is
\begin{eqnarray}
\tilde {\cal G}({\bf k}) = {1\over k^2 + \mu^2}.
\nonumber
%\label{e5.15}
\end{eqnarray}
To obtain the coordinate-space propagator ${\cal G}({\bf r}-{\bf r}')$ we take
the $D$-dimensional Fourier transform of the momentum-space propagator:
\begin{eqnarray}
{\cal G}({\bf r}-{\bf r}') &=& {1\over (2\pi)^D} \int {d^D k\over k^2 + \mu^2}
e^{-i{\bf k}\cdot ({\bf r}-{\bf r}')}\nonumber\\
&=& {1\over (2\pi)^{D/2}} (\mu/ |{\bf r}-{\bf r}'|)^{D-2\over 2}
K_{D-2\over 2}(\mu |{\bf r}-{\bf r}'|).
\label{e5.16}
\end{eqnarray}
The coordinate-space propagator satisfies the Green's function differential
equation
\begin{eqnarray}
(\nabla^2 - \mu^2) {\cal G}({\bf r}-{\bf r}')= \delta^{(D)}({\bf r}-{\bf r}').
\nonumber
%\label{e5.17}
\end{eqnarray}

Let us calculate the amplitude for a particle at ${\bf r}$ to propagate {\sl
anywhere}. We obtain this amplitude by integrating ${\cal G}({\bf r}-{\bf r}')$
with respect to ${\bf r}'$ over all space:
\begin{eqnarray}
\int d^D r' \, {\cal G}({\bf r}-{\bf r}') = {1\over \mu^2}.
\label{e5.18}
\end{eqnarray}

We can now verify that ${\cal G}(r\to r')$ in Eq.~(\ref{e5.13}) is properly
normalized by calculating the amplitude for a particle at radius $r$ to
propagate to {\sl any} radius:
\begin{eqnarray}
\int_0^{\infty} dr'&& {\cal G}(r\to r')\nonumber\\
&& =r^{1-D/2}K_{D-2\over 2}(\mu r)\int_0^r dr'\, (r')^{D/2} I_{D-2\over 2}
(\mu r') +r^{1-D/2}I_{D-2\over 2}(\mu r)\int_r^{\infty} dr'\, (r')^{D/2}
K_{D-2\over 2} (\mu r')\nonumber\\
&& ={r\over \mu^2}\left [ I_{D\over 2}(r)K_{D-2\over 2}(r)+I_{D-2\over
2}(r)K_{D\over 2}(r)\right ]\nonumber\\
&& = {1\over \mu^2},
\nonumber
%\label{e5.19}
\end{eqnarray}
where we have the used the Wronskian identity for modified Bessel functions.
This result agrees with that in Eq.~(\ref{e5.18}).

\subsection{Continuum derivation of spherically symmetric propagator}
\label{ss5.3}
In this subsection we derive the spherically symmetric propagator in
Eq.~(\ref{e5.13}) from the translationally symmetric propagator in 
Eq.~(\ref{e5.16}) by taking an angular average. To obtain the angular
average we let
\begin{eqnarray}
|{\bf r}-{\bf r}'| =\sqrt{r^2+(r')^2-2rr'\cos\theta}.
\nonumber
%\label{e5.20}
\end{eqnarray}
We then expand the modified Bessel function in Eq.~(\ref{e5.16}) as a series in
terms of Gegenbauer polynomials:
\begin{eqnarray}
{K_{\nu}(\mu |{\bf r}-{\bf r}'|)\over |{\bf r}-{\bf r}'|^{\nu}}
=\Gamma(\nu)\left ({1\over 2} \mu^2 rr'\right )^{-\nu}
\sum_{n=0}^{\infty} (n+\nu){\cal C}_n^{(\nu)}(\cos\theta)I_{n+\nu}(\mu r)
K_{n+\nu}(\mu r')\quad (r<r').
\nonumber
%\label{e5.21}
\end{eqnarray}
If we then integrate over the angle $\theta$, only the $n=0$ term
in the series survives and we obtain the result in Eq.~(\ref{e5.13}). The fact
that we obtain the same two-point Green's function directly from our random walk
model supports the validity of the uniform approximation for the probabilities
$P_{\rm out}(n)$ and $P_{\rm in}(n)$ in Eq.\ (\ref{e3.2}).

\section{ACKNOWLEDGMENTS}
\label{s6}
We thank the U.S. Department of Energy for Financial Support. One of us (PNM)
thanks the U.S. Department of Education for additional financial support in the
form of a GANN Predoctoral Fellowship.

\figure{
Comparison between $P_{\rm in}(n)$ in Eq.~(1.2) and the uniform
approximation to $P_{\rm in}(n)$ in Eq.~(3.2) for $D=3$ and $D=5$. Note that the
uniform approximation is exact at $D=1$ and $D=2$.
\label{f1}
}

\begin{table}
\caption{Actual and Predicted Values of $\sum_{n=1}^{\infty} n^k C_{n,t;m}$
for $t = 1000$ and $m = 1$}
\begin{tabular}{lccccc}
$k$&$D$&{Actual}&{Leading Behavior}&{with 1st Correction}&
{with 2nd Correction}\\
\tableline
1&1&$2.622502 \, \times \, 10^{1}$&$2.523133 \, \times \, 10^{1}$&
$2.623133 \, \times \, 10^{1}$&$2.623133 \, \times \, 10^{1}$\\
{}&2&$4.013810 \, \times \, 10^{1}$&$3.963327 \, \times \, 10^{1}$&
$4.013327 \, \times \, 10^{1}$&$4.013823 \, \times \, 10^{1}$\\
{}&3&$5.047526 \, \times \, 10^{1}$&$5.046265 \, \times \, 10^{1}$&
$5.046265 \, \times \, 10^{1}$&$5.047527 \, \times \, 10^{1}$\\
{}&4&$5.987220 \, \times \, 10^{1}$&$5.944991 \, \times \, 10^{1}$&
$5.894991 \, \times \, 10^{1}$&$5.897220 \, \times \, 10^{1}$\\
{}&5&$6.631717 \, \times \, 10^{1}$&$6.728353 \, \times \, 10^{1}$&
$6.628353 \, \times \, 10^{1}$&$6.631718 \, \times \, 10^{1}$\\
\\
2&1&$1.051450 \, \times \, 10^{3}$&$1.000000 \, \times \, 10^{3}$&
$1.050463 \, \times \, 10^{3}$&$1.051463 \, \times \, 10^{3}$\\
{}&2&$2.040138 \, \times \, 10^{3}$&$2.000000 \, \times \, 10^{3}$&
$2.039633 \, \times \, 10^{3}$&$2.040133 \, \times \, 10^{3}$\\
{}&3&$3.001000 \, \times \, 10^{3}$&$3.000000 \, \times \, 10^{3}$&
$3.000000 \, \times \, 10^{3}$&$3.001000 \, \times \, 10^{3}$\\
{}&4&$3.943028 \, \times \, 10^{3}$&$4.000000 \, \times \, 10^{3}$&
$3.940550 \, \times \, 10^{3}$&$3.943050 \, \times \, 10^{3}$\\
{}&5&$4.870366 \, \times \, 10^{3}$&$5.000000 \, \times \, 10^{3}$&
$4.865433 \, \times \, 10^{3}$&$4.870433 \, \times \, 10^{3}$\\
\\
3&1&$5.352671 \, \times \, 10^{4}$&$5.046265 \, \times \, 10^{4}$&
$5.346265 \, \times \, 10^{4}$&$5.352573 \, \times \, 10^{4}$\\
{}&2&$1.219251 \, \times \, 10^{5}$&$1.188998 \, \times \, 10^{5}$&
$1.218998 \, \times \, 10^{5}$&$1.219246 \, \times \, 10^{5}$\\
{}&3&$2.019011 \, \times \, 10^{5}$&$2.018506 \, \times \, 10^{5}$&
$2.018506 \, \times \, 10^{5}$&$2.019011 \, \times \, 10^{5}$\\
{}&4&$2.914616 \, \times \, 10^{5}$&$2.972495 \, \times \, 10^{5}$&
$2.912495 \, \times \, 10^{5}$&$2.914651 \, \times \, 10^{5}$\\
{}&5&$3.892603 \, \times \, 10^{5}$&$4.037012 \, \times \, 10^{5}$&
$3.887012 \, \times \, 10^{5}$&$3.892731 \, \times \, 10^{5}$\\
\\
4&1&$3.205902 \, \times \, 10^{6}$&$3.000000 \, \times \, 10^{6}$&
$3.201851 \, \times \, 10^{6}$&$3.205851 \, \times \, 10^{6}$\\
{}&2&$8.237810 \, \times \, 10^{6}$&$8.000000 \, \times \, 10^{6}$&
$8.237800 \, \times \, 10^{6}$&$8.237800 \, \times \, 10^{6}$\\
{}&3&$1.500000 \, \times \, 10^{7}$&$1.500000 \, \times \, 10^{7}$&
$1.500000 \, \times \, 10^{7}$&$1.500000 \, \times \, 10^{7}$\\
{}&4&$2.342114 \, \times \, 10^{7}$&$2.400000 \, \times \, 10^{7}$&
$2.340550 \, \times \, 10^{7}$&$2.342150 \, \times \, 10^{7}$\\
{}&5&$3.344349 \, \times \, 10^{7}$&$3.500000 \, \times \, 10^{7}$&
$3.338520 \, \times \, 10^{7}$&$3.344520 \, \times \, 10^{7}$\\
\end{tabular}
\label{t1}
\end{table}

\begin{table}
\caption{Actual and Predicted Values of $\sum_{n=1}^{\infty} n^k C_{n,t;m}$
for $t = 1000$ and $D = 2$}
\begin{tabular}{lccccc}
$k$&$m$&{Actual}&{Leading Behavior}&{with 1st Correction}&
{with 2nd Correction}\\
\tableline
1&1&$4.013810 \, \times \, 10^{1}$&$3.963327 \, \times \, 10^{1}$&
$4.013327 \, \times \, 10^{1}$&$4.013823 \, \times \, 10^{1}$\\
{}&3&$4.019752 \, \times \, 10^{1}$&$3.963327 \, \times \, 10^{1}$&
$4.013327 \, \times \, 10^{1}$&$4.019768 \, \times \, 10^{1}$\\
{}&5&$4.033597 \, \times \, 10^{1}$&$3.963327 \, \times \, 10^{1}$&
$4.013327 \, \times \, 10^{1}$&$4.033639 \, \times \, 10^{1}$\\
{}&7&$4.055306 \, \times \, 10^{1}$&$3.963327 \, \times \, 10^{1}$&
$4.013327 \, \times \, 10^{1}$&$4.055438 \, \times \, 10^{1}$\\
{}&9&$4.084814 \, \times \, 10^{1}$&$3.963327 \, \times \, 10^{1}$&
$4.013327 \, \times \, 10^{1}$&$4.085163 \, \times \, 10^{1}$\\
\\
2&1&$2.040138 \, \times \, 10^{3}$&$2.000000 \, \times \, 10^{3}$&
$2.039633 \, \times \, 10^{3}$&$2.040133 \, \times \, 10^{3}$\\
{}&3&$2.046198 \, \times \, 10^{3}$&$2.000000 \, \times \, 10^{3}$&
$2.039633 \, \times \, 10^{3}$&$2.046133 \, \times \, 10^{3}$\\
{}&5&$2.060336 \, \times \, 10^{3}$&$2.000000 \, \times \, 10^{3}$&
$2.039633 \, \times \, 10^{3}$&$2.060133 \, \times \, 10^{3}$\\
{}&7&$2.082553 \, \times \, 10^{3}$&$2.000000 \, \times \, 10^{3}$&
$2.039633 \, \times \, 10^{3}$&$2.082133 \, \times \, 10^{3}$\\
{}&9&$2.112848 \, \times \, 10^{3}$&$2.000000 \, \times \, 10^{3}$&
$2.039633 \, \times \, 10^{3}$&$2.112133 \, \times \, 10^{3}$\\
\end{tabular}
\label{t2}
\end{table}


\begin{references}
\bibitem{wow} The result in Eq.~(4.5) at $m=1$ can be rederived using the
continued-fraction methods of Ref.~[1]. We simply substitute
$$Q_n=P_{\rm out}(n)P_{\rm in}(n+1)={n(n+D-2)\over (2n+D-3)(2n+D-1)}$$
into Eq.~(3.21) of Ref.~[1]. We obtain a continued-fraction representation
for $\Pi_1(D)$ which can be simplified to read
$$\Pi_1(D)=1/(D+1-2(D)/(D+3-3(D+1)/(D+5-4(D+2)/(D+7-5(D+3)/(\cdots))))).$$
If this continued fraction is truncated before the factors $(D)$, $(D+1)$,
$(D+2)$, $(D+3)$, and so on, each successive truncation gives exactly
$\Pi_1(D)={1\over D-1}$.
\end{references}
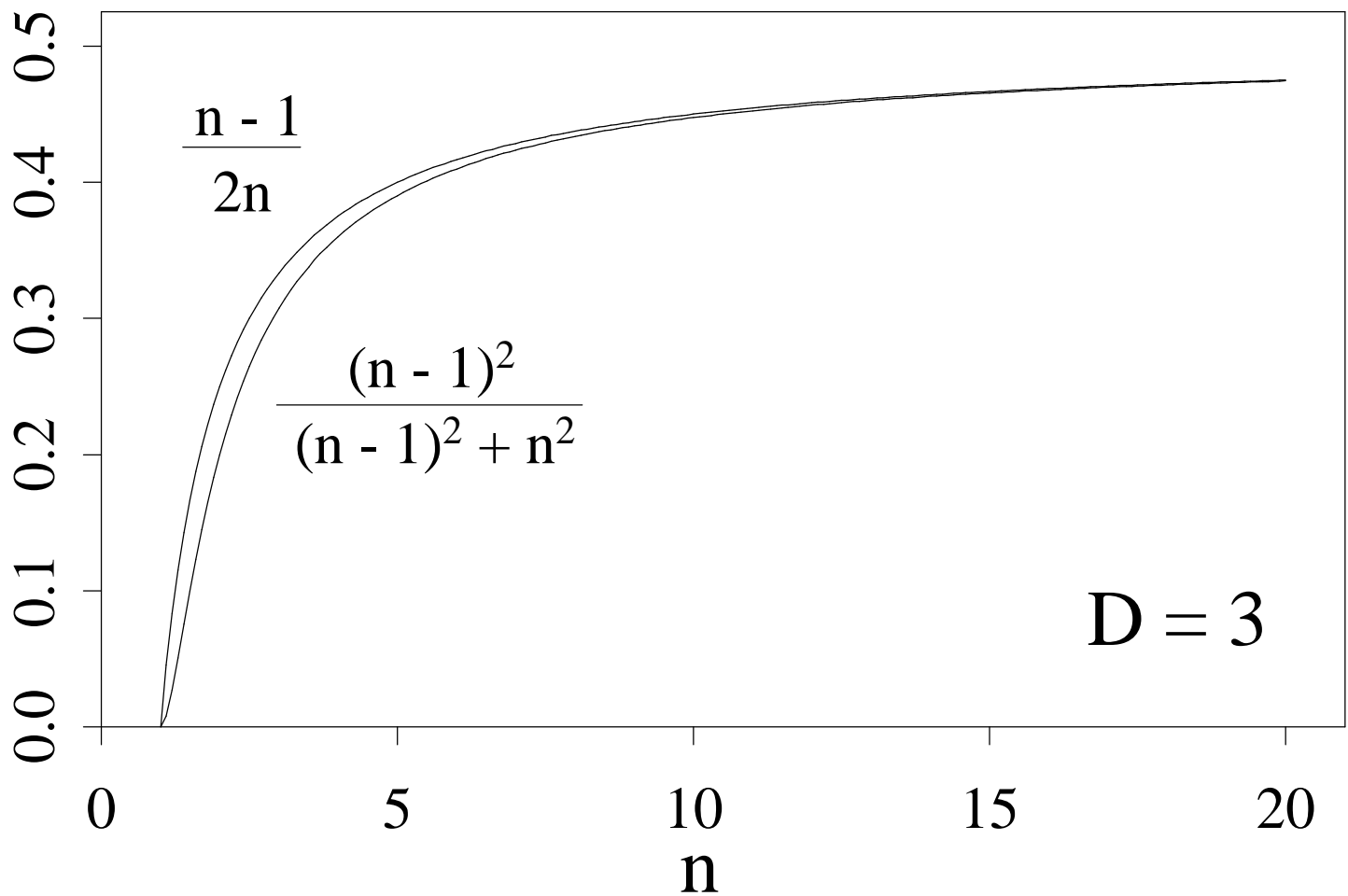
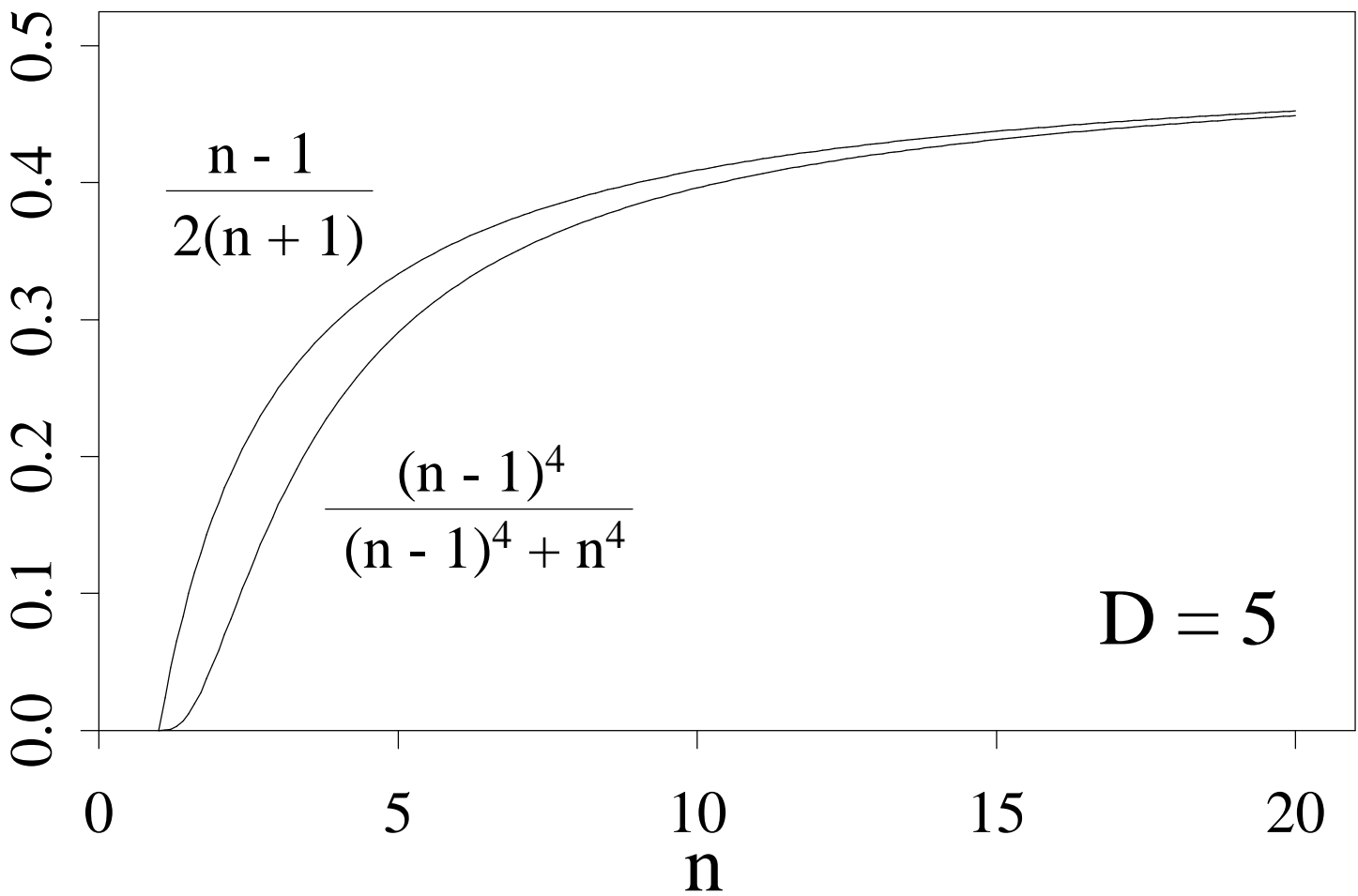
\end{document}